\documentclass[preprint,prl,showpacs,aps,superscriptaddress]{revtex4-1}
\usepackage{hyperref}
\usepackage{txfonts}
\usepackage{graphicx}
 
\usepackage[utf8]{inputenc}

\usepackage{xspace}
\newcommand{\lcvo}{LiCuVO\ensuremath{_4}\xspace}
\newcommand{\tn}{\ensuremath{T_\mathrm{N}}\xspace}
\newcommand{\hn}{\ensuremath{H_\mathrm{c}}\xspace}

\newcommand{\aaxis}{\ensuremath{a}\xspace}
\newcommand{\baxis}{\ensuremath{b}\xspace}
\newcommand{\caxis}{\ensuremath{c}\xspace} 
\renewcommand\vec{\mathbf}
\usepackage{epstopdf}
\begin{document}

\title{Observation of chiral solitons in the quantum spin liquid phase of \lcvo} 
\date{\today} 

\author{Christoph P. \surname{Grams}}
\email{grams@ph2.uni-koeln.de}
\affiliation{University of Cologne, Institute of Physics II, Z\"ulpicher Str. 77, 50937 Cologne, Germany}
\author{Daniel \surname{Br\"uning}}
\affiliation{University of Cologne, Institute of Physics II, Z\"ulpicher Str. 77, 50937 Cologne, Germany}
\author{Severin \surname{Kopatz}}
\affiliation{University of Cologne, Institute of Physics II, Z\"ulpicher Str. 77, 50937 Cologne, Germany}
\author{Thomas \surname{Lorenz}}
\affiliation{University of Cologne, Institute of Physics II, Z\"ulpicher Str. 77, 50937 Cologne, Germany}
\author{Petra \surname{Becker}}
\affiliation{University of Cologne, Institute of Geology and Mineralogy, Section Crystallography, Z\"ulpicher Str. 49b, 50674 Cologne, Germany}
\author{Ladislav \surname{Bohatý}}
\affiliation{University of Cologne, Institute of Geology and Mineralogy, Section Crystallography, Z\"ulpicher Str. 49b, 50674 Cologne, Germany}
\author{Joachim \surname{Hemberger}}
\affiliation{University of Cologne, Institute of Physics II, Z\"ulpicher Str. 77, 50937 Cologne, Germany}

\begin{abstract}
    Quantum spin liquids represent a magnetic ground state arising in the presence of strong quantum fluctuations that preclude ordering down to zero temperature and leave clear fingerprints in the excitation spectra.
    While theory bears a variety of possible quantum spin liquid phases their experimental realization is still scarce.
    Here, we report the first experimental evidence of a vector-chiral quantum spin liquid state in the $S=1/2$ spin chain compound \lcvo from measurements 
    of the complex permittivity $\varepsilon^*$ in the GHz range.  
    In zero magnetic field our results show short-lived thermally activated chiral fluctuations above the multiferroic phase transition at $\tn=2.4$\,K with divergent life-times when approaching \tn.
    In $\varepsilon^*$ this fluctuation dynamics are seen as the slowing down of a relaxation with a critical dynamical exponent $\nu_\xi z \approx 1.3$ 
    in agreement with mean-field predictions.
    When using a magnetic field to suppress \tn towards 0\,K the influence of quantum fluctuations is increased until they condense into the chrial spin liquid phase below 400\,mK. 
    Within this phase we measure a nearly-gapless chiral soliton excitation with a tiny energy gap of $E_\mathrm{SE}\approx14.1$\,\(\mu\)eV.
\end{abstract}

\maketitle
 
\section{Introduction}  
In solid state physics unique phenomena emerge when a long-range order is suppressed by an external parameter towards lower temperatures where quantum fluctuations dominate. 
This suppression often reveals a 'dome'-like new phase at the bottom of a conical cross-over region and a significant change of the critical dynamics~\cite{Sachdev2011}.
Famous examples for this are superconductivity~\cite{Shen2008,Rischau2017},  heavy fermion systems~\cite{Gegenwart2008,Si2010}, and quantum spin liquids~\cite{Savary2017}.
To investigate the influence of quantum fluctuation on the critical dynamics, systems with low dimensionality are a good starting point as they ideally do not order even at $T\to 0$\,K.
Experimentally, purely 1D systems are, of course, not available, but systems where the magnetic ions are located on structures with lower dimensionality, e.g. spin-chain compounds, have been shown to undergo magnetic-field driven phase transitions at very low temperatures.
As this quantum phase transitions are driven by quantum fluctuations and not by temperature unusual dynamics are expected to emerge in their proximity.
The influence of quantum fluctuations is even stronger when the systems have almost isotropic spin $S=1/2$ in so-called Heisenberg spin chains.
Not only are these systems interesting in themselves, but by introducing vector spin chirality, e.g. via Dzyaloshinskii-Moriya interaction, this model can - depending on the details of the chosen interaction - be mapped onto a quantum wire or a 1D superconducting Josephson junction array~\cite{Gangadharaiah2008,Garate2010}.
An alternative way to introduce vector chirality in such a spin chain is frustration due to competing nearest and next nearest neighbor interactions, $J_1$ and $J_2$, in the presence of a small easy-plane anisotropy~\cite{Sato2007,Furukawa2008}. 
Apart from being chiral, these phases can be magnetoelectric multiferroic~\cite{Spaldin2005, Khomskii2006} as has been observed in many compounds~\cite{Khomskii2009, Fiebig2016}. 
The coupling between magnetization $M$ and polarization $P$ has potential applications for data storage~\cite{Scott2007,Shang2018} and, most important for this work, it allows to observe the dynamics of the magnetic system via measurements of the polarization dynamics~\cite{Niermann2015,Grams2019}. 
In particular, in the case of magnetoelectric multiferroics an electromagnon~\cite{Pimenov2006,ValdesAguilar2009}, i.e., collective magnetic excitations that also carry electric polarization, can be observed.
The corresponding fluctuation dynamics show critical slowing down at the multiferroic phase transition~\cite{Niermann2014}.
This phase transition can be tuned towards $T\to0$\,K in an external magnetic field.
Close to this quantum critical point a change in the fluctuation dynamics can be observed via dielectric spectroscopy~\cite{Kim2014a}.

\begin{figure} 
  \centering
  \includegraphics[width=\columnwidth]{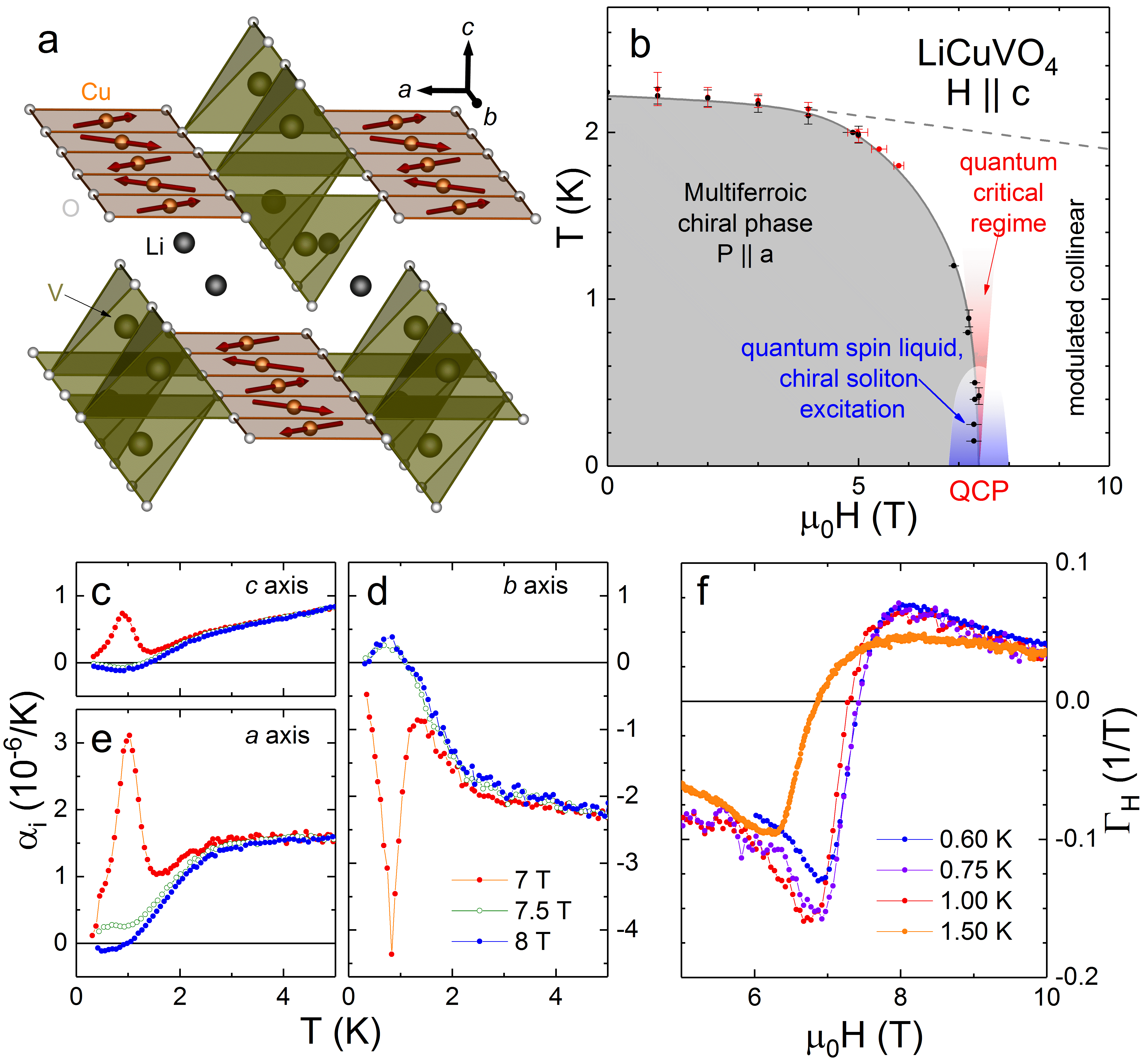}
  \caption{
    \textbf{Introduction of LiCuVO$_\mathbf{4}$.}
    \textbf{a} A fraction of the crystal structure of \lcvo (data from~\cite{Enderle2005}), the ordered magnetic moments of Cu in the multiferroic phase are indicated.
    \textbf{b} A sketch of the $H$ vs. $T$ phase diagram (data from \cite{Grams2019}).
    \textbf{c}-\textbf{e}, The thermal expansion coefficients $\alpha_i$ for all 3 crystallographic axes change sign upon crossing the critical magnetic field $\mu_0\hn\simeq7.4$\,T and, \textbf{f}, an analogous sign change occurs in the magnetic Gr\"uneisen parameter.
  }
  \label{fig:intro}
\end{figure}
 
\section{\lcvo} 

In this work we study \lcvo with a distorted inverse spinel-type structure~\cite{Lafontaine1989}, see Figure~\ref{fig:intro}a, where competing ferromagnetic $J_1<0$ and antiferromagnetic $J_2>0$ interactions~\cite{Enderle2005} lead to frustration.
Here, a low-dimensional spin system is realized by Cu$^{2+}$ ions with $S=1/2$ that form 1D chains along the \baxis axis that are separated in \aaxis direction by VO$_4$ tetrahedra and by Li along the \caxis direction.
Cycloidal N\'eel ordering of the spins within the $ab$ plane with an incommensurate propagation vector $\vec{k} =(0, 0.532, 0)$~\cite{Gibson2004} is observed below $\tn=2.4$\,K due to a weak interchain coupling $J_\mathrm{i.c.}$, as sketched in Figure~\ref{fig:intro}b.
In the phase with this cycloidal order \lcvo is multiferroic with an electric polarization $\vec{P} \propto \vec{k} \times (\vec{S}_i \times \vec{S}_{i+1})$ parallel to the \aaxis axis.
By applying a magnetic field along the \caxis axis the 3D order can be weakened and \tn is continuously suppressed down to 0\,K at $\mu_0\hn\approx 7.4$\,T~\cite{Naito2007,Schrettle2008,Grams2019}.

As is shown in Fig.~\ref{fig:intro}c-e, the transition into the multiferroic phase causes sharp anomalies in the uniaxial thermal expansion coefficients $\alpha_i = 1/L \, \partial L / \partial T$ ($i$ denotes the orthorhombic axes \aaxis, \baxis, \caxis). 
Upon increasing the magnetic field $H||\caxis$ above \hn, these anomalies vanish and $\alpha_i$ changes sign for each axis $i$~\footnote{For $i=a$, a small anomaly remains present for $\mu_0 H = 7.5$~T and the sign changes of $\alpha_a$ occurs above 7.5\,T. 
This probably arises either from a stabilization of the multiferroic phase for uniaxial pressure along \aaxis and/or from a slight misorientation of the magnetic field.}.
These sign changes arise from sign changes of the corresponding Gr\"uneisen ratios $\Gamma_i= \alpha_i/C_p$ (with the specific heat $C_p$), which are characteristic signatures of pressure-dependent quantum phase transitions~\cite{Zhu2003,Garst2005} as observed in various quasi-1D quantum spin materials~\cite{Lorenz2007, Lorenz2008, Breunig2013, Niesen2013, Breunig2017, Wang2018}.
As shown in Fig.~\ref{fig:intro}f, a sign change also occurs in the magnetic-field dependent Gr\"uneisen paramater $\Gamma_H=1/T \, \partial T / \partial (\mu_0H)$, which is obtained from the ratio of the magnetocaloric effect and $C_p$~\cite{Breunig2017, Wang2018}.
The phase boundary of \lcvo is very steep close to \hn, see Fig.~\ref{fig:intro}b,  which experimentally makes the $H$-dependent measurements of $\Gamma_H$ more reliable than the $T$-dependent determination of $\Gamma$. 

When increasing the magnetic field above \hn at low temperatures the system enters a collinear spin-modulated phase\cite{Buttgen2007} and vector-chiral correlations were reported to exist already for $T>\tn$~\cite{Ruff2019}. 
Above 41\,T \lcvo enters a spin-nematic phase, and a saturated phase above 44\,T~\cite{Svistov2011,Mourigal2012}, both of which are well beyond the magnetic field range used in this work.

\begin{figure*}
  \centering
  \includegraphics[width=\columnwidth]{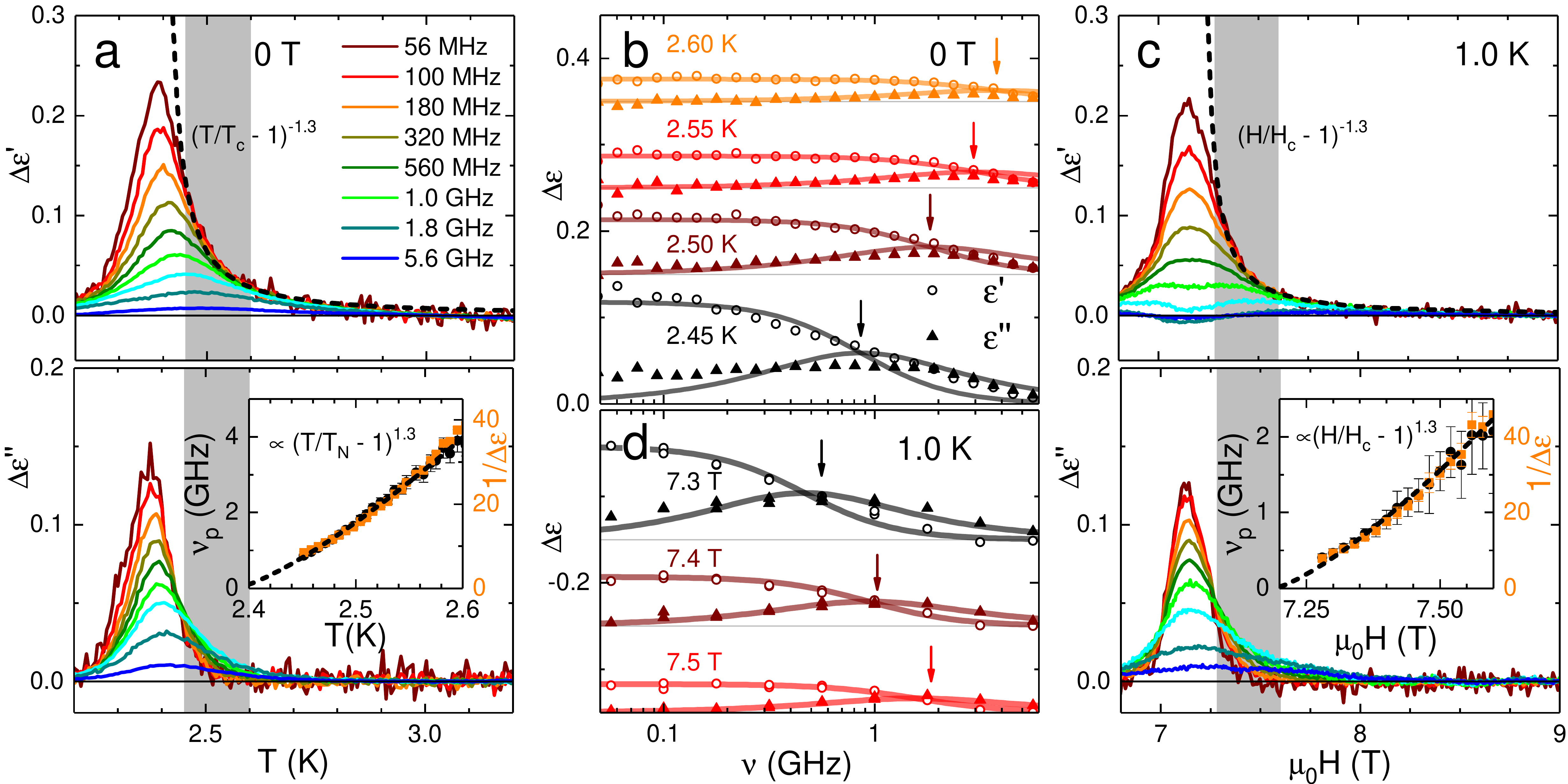} 
  \caption{
    \textbf{\textit{T-} and \textit{H-} dependent permittivity.}
    \textbf{a}, shows the real (upper panel) and imaginary part (lower panel) of the complex permittivity in the frequency range of 56\,MHz - 5.6\,GHz in zero magnetic field, the fit of the critical dynamical exponent is shown in the inset. 
    In \textbf{b}, the measured permittivity data (circles for $\Delta\varepsilon'$ and triangles for $\Delta\varepsilon''$), are shown together with the fitted Debye relaxation as solid lines.
    The curves are offset for enhanced visibility and the peak positions in the dielectric loss marked by arrows.
    The reduction of $\nu_\mathrm{p}$ when closing in on \tn demonstrates critical slowing down.
    \textbf{c} shows the $H$ dependent complex permittivity at 1.0\,K with the fit of the critical exponent in the inset. 
    Spectra with fits at 1.0\,K in \textbf{d} show the slowing down in magnetic field.
    } 
 \label{fig:data} 
\end{figure*}

Here, we report the first experimental evidence of a chiral quantum spin liquid state in \lcvo.
This evidence is based on the analysis of the $T$-, $H$-, and $\nu$-dependence of the complex permittivity $\varepsilon^*$ of \lcvo close to the multiferroic phase transition down to 0.025\,K with frequencies in the GHz range. 
Our results show classical critical slowing down at the multiferroic phase transition with a critical dynamical exponent $\nu_\xi z \approx 1.3$ above 1.0\,K in agreement with mean-field predictions~\cite{Niermann2015}.
For an external magnetic field the transition is shifted towards low temperatures.
In this regime we observe quantum critical slowing down of the fluctuations that condense into a chiral quantum spin liquid state with a nearly gapless chiral soliton excitation.
From direct measurements we find an energy gap of $E_\mathrm{SE}\approx14.1$\,\(\mu\)eV for this excitation.

 
In Figure~\ref{fig:data} measurements of the permittivity in the frequency range from 56\,MHz to 5.6\,GHz are presented.
Figure~\ref{fig:data}a shows the temperature dependence in zero magnetic field with each curve corresponding to a fixed frequency. The upper panel shows the real part of the permittivity, the corresponding imaginary part, the dielectric loss, is plotted in the lower panel.
Here and in all results, the background of the permittivity away from the phase transition was subtracted as it is not caused by cycloidal spin-fluctuations but regular phonon contributions that are always present in measurements of the permittivity. 
On approaching the multiferroic phase transition fluctuations of this phase appear already above \tn and their size and life-time diverge for $T \to \tn$.
In this regime dispersion is observed, as the high frequencies break away from the quasi-static response that is given by the envelope of $\Delta\varepsilon'$ (dashed line in Figure~\ref{fig:data}a) and is accompanied by corresponding contributions in the dielectric loss at temperatures above \tn.
These characteristic features of critical slowing down are expected in both ferroelectric as well as multiferroic materials~\cite{Niermann2015}. 

A different view of a selection of the same data is shown in Figure~\ref{fig:data}b, where the complex permittivity is shown for different temperatures as both, $\varepsilon'(\nu)$ (circles) and $\varepsilon''(\nu)$ (triangles).
In this plot the dispersion discussed above causes a step in $\varepsilon'(\nu)$ that is accompanied by a peak in $\varepsilon''(\nu)$. 
This behavior results from the phonon mode's angular eigenfrequency $\omega_0$ that vanishes close to \tn, while the damping $\Gamma > \omega_0$ remains finite.
Once the softening mode is observed in the GHz regime it can be described as a Debye relaxation with a relaxation time $\tau \approx \Gamma/\omega_0^2$ in measurements of its dynamics,
\begin{equation}
  \varepsilon_\mathrm{EM}(\omega) = \frac{\Delta\varepsilon_\mathrm{s}}{1+i\omega\tau}.
  \label{eq:relaxation}
\end{equation}
Here, $\Delta\varepsilon_\mathrm{s}$ denotes the change in the static permittivity, i.e. the height of the observed step in $\varepsilon'$ at low frequencies and the angular frequency $\omega=2\pi\nu$ is connected to the frequency $\nu$ used in the experiments.
The fits of the Debye relaxation to the complex permittivity are also shown in Figure~\ref{fig:data}b as lines for both, $\varepsilon'(\nu)$ and $\varepsilon''(\nu)$.
Here, the arrows mark the position $\nu_\mathrm{p}=1/2\pi\tau$ of the maxima in the dielectric loss, and their shift towards lower frequencies upon cooling demonstrates the relaxation's critical slowing down close to the phase transition.
Additional information on the dynamics can be gained from the height of the step, $\Delta\varepsilon_\mathrm{s}$.
According to the Lyddane-Sachs-Teller relation~\cite{Blinc1974} there is a connection between $\omega_0$ and the critical increase in the static permittivity, $\Delta\varepsilon_\mathrm{s} \propto \omega_0^{-2}$.
As $\omega_0^2 \approx \Gamma/\tau$ also in the case of high damping $\Delta\varepsilon_\mathrm{s}$ is connected to the relaxation time by $\Delta\varepsilon_\mathrm{s} \propto \tau$.

To quantify the dynamics above the phase transition we utilize critical dynamical scaling that relates the relaxation time $\tau \propto \xi^z$ to the correlation length $\xi$ via the dynamical exponent $z$.
Close to the phase transition the correlation length $\xi$ diverges, $\xi(T) \propto |T/\tn -1|^{-\nu_{\xi,T}}$, with the positive correlation length exponent $\nu_\xi$.
Consequently, for the $T$ dependence of the relaxation time we expect $\tau(T) \propto |T/\tn-1|^{-\nu_{\xi,T}z}$ and, at the same time, $\Delta\varepsilon_\mathrm{s} \propto |T/\tn -1|^{-\nu_{\xi,T} z}$.
In the inset of Figure~\ref{fig:data}a we show that a critical exponent  $\nu_\xi z \approx 1.3$ simultaneously describes $\nu_\mathrm{p}=1/2\pi\tau$ and $1/\Delta\varepsilon_\mathrm{s}$, in agreement with the expectation of $\nu_{\xi,T} z \approx 1.272$ for materials of the 3D-Ising universality class~\cite{Niermann2015}.


Measurements of $\varepsilon^*(H)$ were performed at constant temperatures. 
Figure \ref{fig:data}c shows a measurement at 1.0\,K that looks very similar to $\varepsilon^*(T)$ shown in Figure \ref{fig:data}a, the only difference being the directions of crossing the phase transition.
Compared to the zero magnetic field measurement the overall size of the feature is slightly reduced.
This reduction is caused by the phase transition's shifts towards higher magnetic fields, thus canting the spins further into the direction of the magnetic field and, thereby, reducing the contribution of $(\vec{S}_i \times \vec{S}_{i+1})$ to the polarization. 
The spectra in Figure~\ref{fig:data}d demonstrate that critical slowing down is also observed for $H$-dependent measurements at constant temperature.

We again evaluate the dynamical critical exponent, in this case for its magnetic-field dependence: $\tau(H) \propto |H/\hn -1|^{-\nu_{\xi,H} z}$.
The result is shown as a dashed line in Figure~\ref{fig:data}c, and the fit to $\nu_\mathrm{p}$ and $1/\Delta\varepsilon_\mathrm{s}$ in the inset describes the data with the same critical exponent, $\nu_{\xi,H} z \approx 1.3$, as found for the zero-field data in agreement with the expected mean-field behavior~\cite{Niermann2015}.

\subsection{Nearly gapless excitation} 

\begin{figure*} 
  \centering
  \includegraphics[width=0.7\columnwidth, trim=0 0 10cm 0]{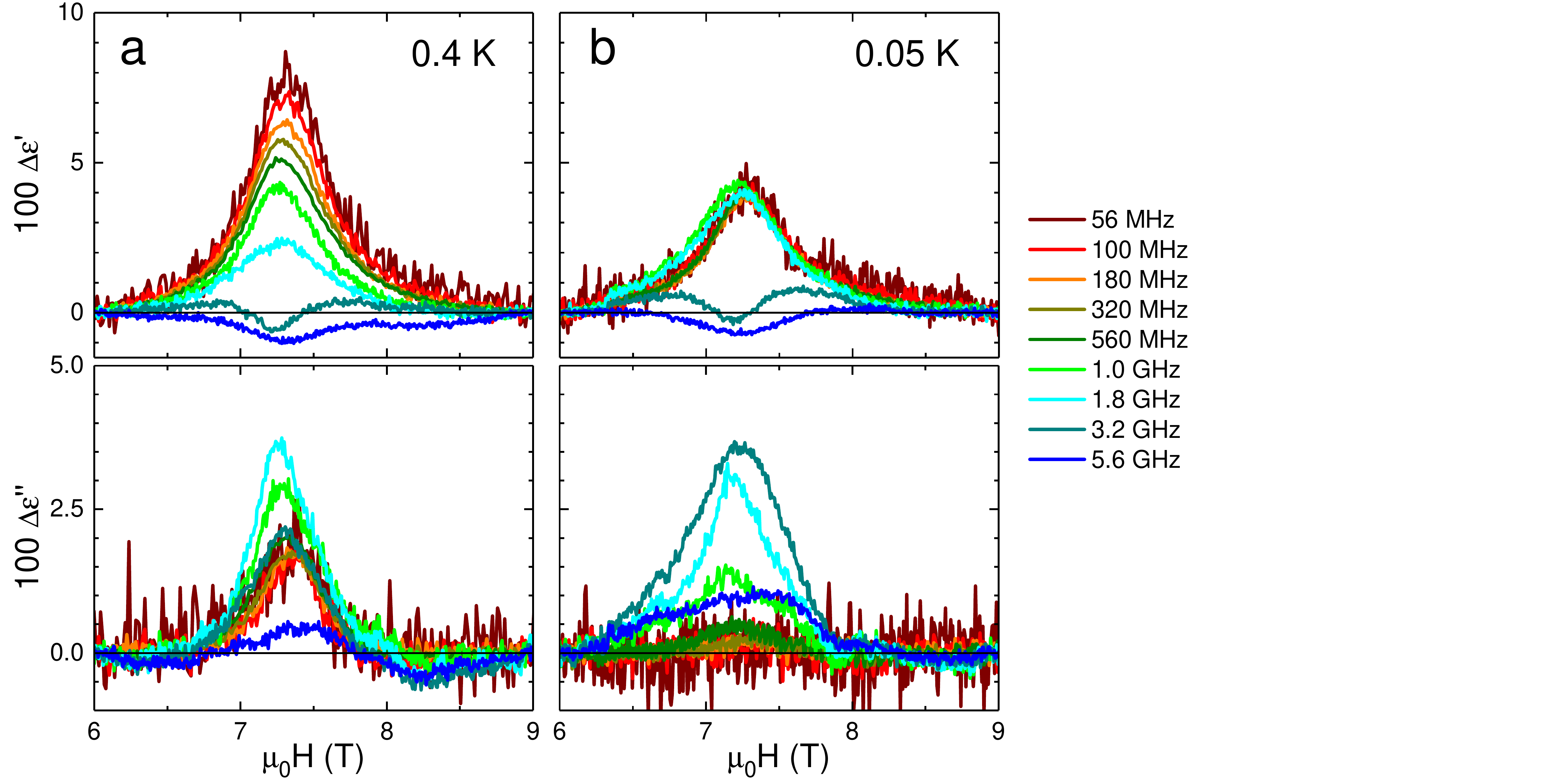}
  \caption{
    \textbf{Low temperature dynamics.}
    The real (upper panel) and imaginary part (lower panel) of the complex permittivity in the frequency range of 56\,MHz - 5.6\,GHz are shown for 0.4\,K in \textbf{a} and 0.05\,K in \textbf{b}.
    } 
 \label{fig:lowT}
\end{figure*} 

Following the phase transition towards 0.4\,K in Figure~\ref{fig:lowT}a we still see the typical dispersion feature in $\varepsilon'(H)$ even though its height is reduced due to the further canting of the spins in the magnetic field.
The first indication of a fundamental change in the dynamics at the phase transition is seen in the dielectric loss in the lower panel.
While we still observe dispersion at the phase transition the contributions from lower frequencies at the phase transition are suppressed, the highest signal is observed at around 1\,GHz.

At even lower temperatures, exemplary shown for 0.05\,K in Figure~\ref{fig:lowT}b and in more detail presented in the supplementary material, the dispersion in $\Delta\varepsilon'(H)$ vanishes for frequencies below 1\,GHz and, at the same time, the dielectric loss shows no contributions in this frequency range.
As the highest dielectric loss is measured at the same frequency for all magnetic fields the observed dynamics are, within the experimental accuracy, independent of $H$.

\begin{figure}
 \centering 
 \includegraphics[width=0.9\columnwidth]{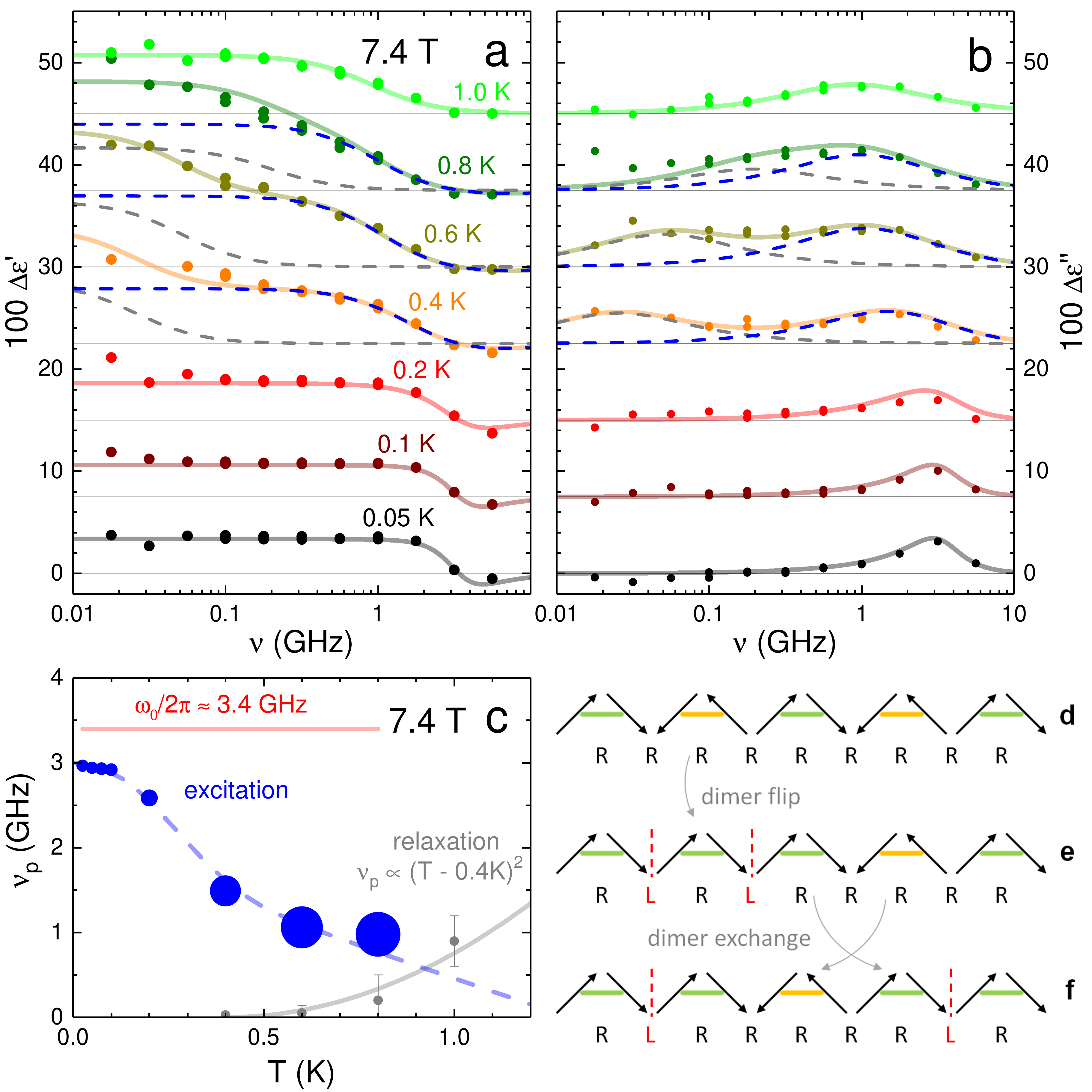}
 \caption{ 
   \textbf{The emergence of chiral solitons.}
   Measured spectra (points) at $H\approx\hn$ with fits of Equation~(\ref{eq:excitation}) (lines) are shown in \textbf{a} (real part) and \textbf{b} (imaginary part).
   The fitted peak frequencies of the maxima in $\varepsilon''$ are shown in \textbf{c} for both the relaxation and the emergent excitation; the size of the points indicates the change in damping for the excitation.
   \textbf{d} The pristine spin chain with uniform clockwise vector handedness. 
   Dimers are marked by colored dashes to differentiate between the alternating species that make up the spin chain.
   \textbf{e} By flipping both spins in a dimer it transforms into the other species and two chiral solitons are created.
   This process is slightly gapped due to the presence of interchain coupling in \lcvo.
   \textbf{f} To move the solitons through the spin chain two neighboring dimers are exchanged and, thus, the soliton moves without creating additional defects in the handedness of the spin chain. 
 }
 \label{fig:spectra}
\end{figure}

To examine the change in the fluctuation dynamics in more detail we show the real and imaginary part of $\varepsilon^*$ as a function of frequency $\nu$ at several temperatures and constant magnetic field $\mu_0 H = 7.4$\,T, i.e. very close to \hn, in Figure~\ref{fig:spectra}a and b.
As shown before, a single relaxation process is sufficient to describe the data at 1.0\,K.
Below 1.0\,K a second process emerges in the measurement and an additional Lorentz oscillator has to be added:
\begin{equation}
  \varepsilon^*(\omega) = \varepsilon_\mathrm{EM}(\omega) + \frac{\omega_\mathrm{p}^2}{\omega_0^2 - \omega^2 +i\omega\Gamma}.
  \label{eq:excitation}
\end{equation}
Here $\omega_\mathrm{p}$ denotes the plasma frequency, $\omega_0$ the undamped eigenfrequency, and $\Gamma$ the damping.
The results of the fits are shown as lines in both panels and are in good agreement with the data, the two components are also shown as gray (relaxation) and blue (excitation) dashed lines.

First, we focus on the temperature dependence of the relaxation shown as dashed gray lines in Figure~\ref{fig:spectra}a and b.
During cooling the relaxation is slowing down until it cannot be observed below 0.4\,K.
The peak frequency $\nu_\mathrm{p}$ from the fits is shown as gray dots in Figure~\ref{fig:spectra}c.
When measuring the permittivity in multiferroic materials this dependence has been shown to be modified to  $\tau \propto 1/T^2$~\cite{Kim2014a}, and our results are in agreement with this critical exponent (solid line).

The second contribution, shown in the intermediate temperature range as dashed blue lines in Figure~\ref{fig:spectra}a and b, on the other hand, does not show slowing down upon cooling, but instead the peak shifts to higher frequencies.
In Figure~\ref{fig:spectra}c $\nu_\mathrm{p}$ of this contribution is shown as blue dots.
This apparent 'speeding up'~\cite{Grams2014} can be explained by a reduced damping $\Gamma$ with decreasing $T$, indicated by the size of the dots in the figure, while the eigenfrequency $\omega_0$ is temperature independent. 
While the observed damping at "high" temperatures is of similar magnitude as the eigenfrequency it's reduction upon cooling demonstrates that this contribution indeed has excitational character.

In a theoretical work by Furukawa et al.~\cite{Furukawa2008} it has been shown that for $S=1/2$ systems quantum fluctuations can lead to the emergence of a spin-liquid state in which the spins dimerize.
The idea that the authors proposed works as follows: starting from a uniform spin chain in Figure~\ref{fig:spectra}d the suppression of long range order leads to dimerization, marked by green and yellow dashes in the figure.
In the beginning the vector handedness of the spiral is uniform, here clockwise (right handed).
By flipping both spins in a dimer two 'domain walls' in the cycloidal 1D spin chain are created.
These domain walls can be considered as chiral solitons due to the change in the local handedness of the spin spiral from clockwise (right handed) to counter clockwise (left handed), as marked in Figure~\ref{fig:spectra}e.
Due to this change in the handedness the solitons also carry an electric dipole moment that locally points against the ferroelectric background of the uniform spin chain and are thus visible by dielectric spectroscopy.
Once the chiral solitons are created they can move through the spin chain when two neighboring dimers exchange their positions (Figure~\ref{fig:spectra}f).
These excitations are slightly gapped due to the interchain coupling in \lcvo and are driven by external electric fields.
Our results show that the gap of the soliton excitation is indeed very small, $\omega_0 / 2\pi \approx 3.4$\,GHz which corresponds to a gap energy of $E_\mathrm{SE} = 14.1$\,$\mu$eV.

\subsection{Conclusion}
Our measurements of the complex permittivity close to the multiferroic phase transition in \lcvo demonstrate the changing polarization dynamics in the presence of increasing quantum fluctuations.
At temperatures above 1.0\,K we see the mean-field behavior expected for the critical dynamics at a multiferroic phase transition: critical slowing down 
for both, the temperature and magnetic-field dependent measurements.
When \tn is suppressed further the critical slowing down of thermal fluctuations is insufficient to describe the dynamics.
Instead, thermal and quantum fluctuations compete until the later condense into a vector-chiral quantum spin liquid phase.
In this novel phase we measure a contribution to the permittivity spectra that, despite being in the GHz regime, does not show significant damping.
This low-energy excitation with a tiny gap of $E_\mathrm{SE}\approx14.1$\,\(\mu\)eV is consistent with the predicted chiral soliton excitation in \lcvo~\cite{Furukawa2008}.

%

\section*{Acknowledgment}

This work is funded through the Institutional Strategy of the University of Cologne within the German Excellence Initiative and the Deutsche Forschungsgemeinschaft (DFG, German Research Foundation) - Projektnummer 277146847 - SFB 1238 (A02, B01, and B02).
The authors thank M. Grüninger, C. Kollath, and M. Garst for fruitful discussions.
\appendix
\section{}

\section{Methods}

Single crystals of \lcvo were grown from LiVO$_3$ flux in the temperature range between 910\,K and 853\,K with an applied cooling rate of 0.1\,K/hour.
To observe the fluctuation dynamics at the edge of the multiferroic phase we mounted a \lcvo crystal on a microstrip sample holder for dielectric spectroscopy measurements in the GHz range.
The measurements where done with different network analyzers (Agilent PNA-X and Rohde \& Schwarz ZVB4 and ZNB8) in two cryostats, a top-loading dilution refrigerator (Oxford Instruments KELVINOX) and a Quantum Design PPMS.
Thermal expansion was measured in a home-built capacitance dilatometer with an ac capacitance bridge (Andeen-Hagerling AH2550).
The uniaxial thermal expansion coefficient was obtained numerically via $\alpha_i = 1/L_i \partial \Delta L_i/\partial T$.
The magnetic Gr\"uneisen parameter $\Gamma_H$ was measured using a home-built calorimeter with a technique explained in~\cite{Breunig2017}. 

\bibliographystyle{apsrev4-1} 
\bibliography{LiCuVO4-HFPaper}

\onecolumngrid
\newpage
\section{Supplementary Information}

\begin{figure*}[h]
  \centering
  \includegraphics[width=0.44\columnwidth]{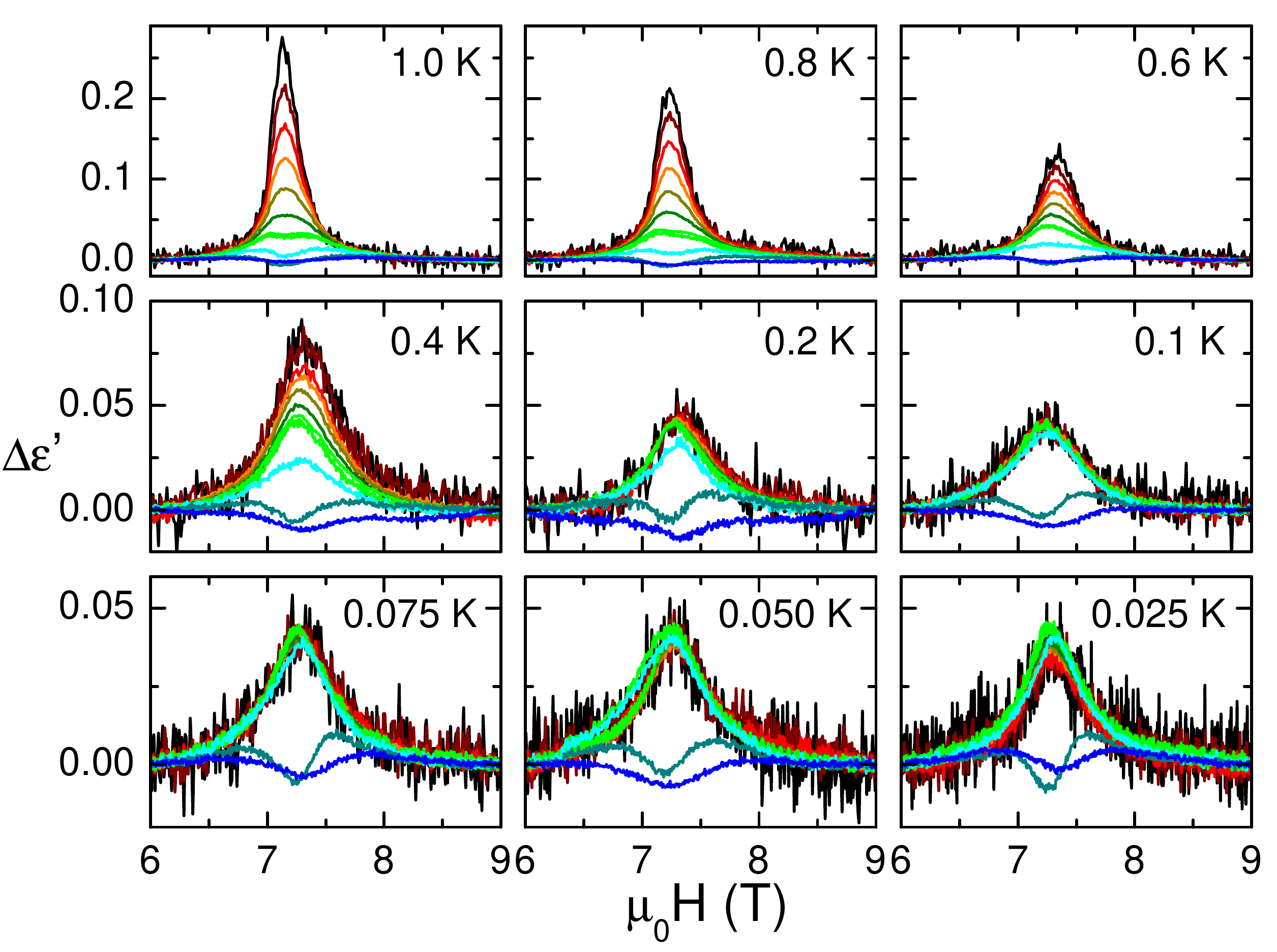}
  \includegraphics[width=0.1\columnwidth, trim= 0mm -60mm 0mm 0mm]{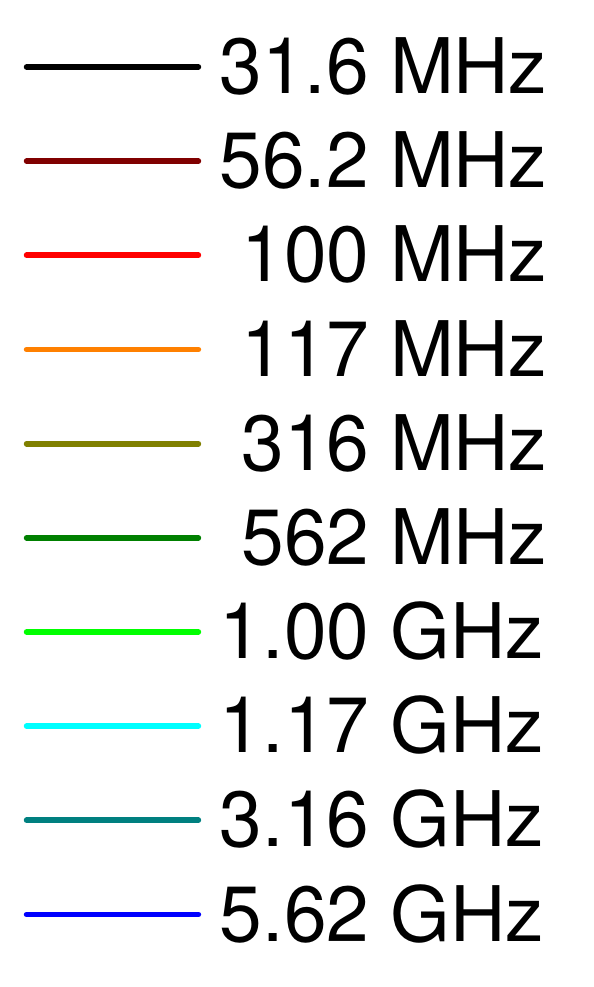} 
  \includegraphics[width=0.44\columnwidth]{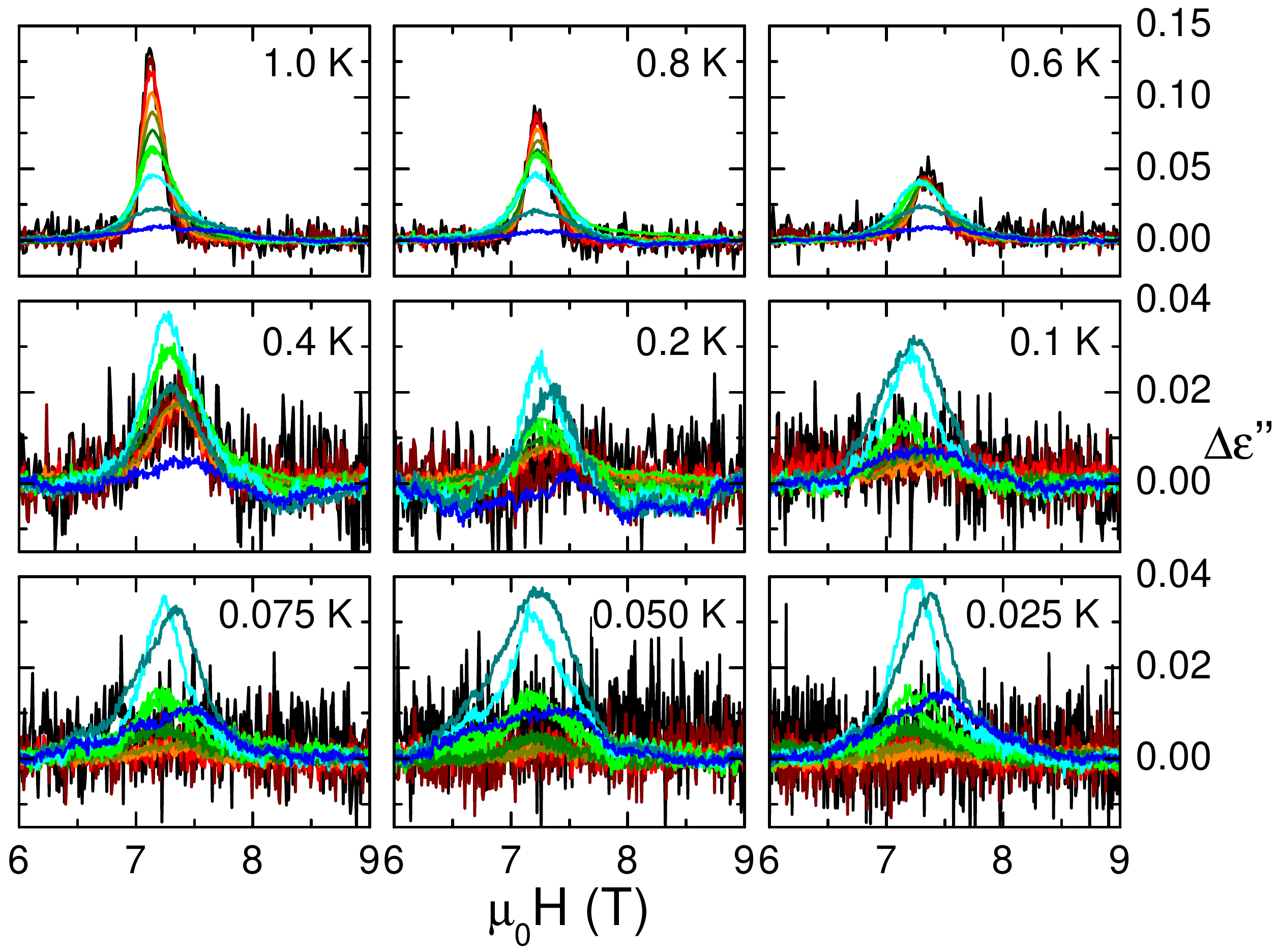} 
  \caption{
    $\Delta\varepsilon^*$ as function of the magnetic field for different frequencies and temperature. 
    Contributions of faster polarization effects have been subtracted. 
    \textbf{a}, shows the real part $\Delta\varepsilon'$ where dispersion with a dispersion minimum at high frequencies can be see at the higher temperature.  
    Below 0.4\,K the dispersion at the peak vanishes for frequencies up to 1\,GHz (green). 
    Above 1\,GHz the peak is changed to a minimum. 
    The dielectric loss $\Delta\varepsilon''$ in \textbf{b} also changes from a dispersive maximum at higher temperatures to a peak between 2 and 3\,GHz.}
 \label{fig:SI-fig1}
\end{figure*}

\end{document}